\documentclass[journal=jacsat,manuscript=article]{achemso}

\usepackage[version=3]{mhchem} % Formula subscripts using \ce{}
\usepackage{breqn}
\usepackage{xcolor}
\usepackage{graphicx}
\usepackage{float}
\usepackage[export]{adjustbox}
\usepackage{balance}

\makeatletter

\newcommand*\datastatementname{Data availability}
\makeatother

\author{Philipp Hönicke}
\affiliation{Helmholtz-Zentrum Berlin (HZB), Hahn-Meitner-Platz 1, 14109 Berlin, Germany}
\alsoaffiliation{Physikalisch-Technische Bundesanstalt (PTB), Abbestr. 2-12, 10587 Berlin, Germany}

\author{Yves Kayser}
\affiliation{Max Planck Institute for Chemical Energy Conversion, Stiftstr. 34-36, 45470 Mülheim an der Ruhr, Germany}

\author{Pouya Partovi-Azar}
\affiliation{Institute of Chemistry, Martin Luther University Halle-Wittenberg, Von-Danckelmann-Platz 4, 06120 Halle (Saale), Germany}
\email{pouya.partovi-azar@chemie.uni-halle.de}

\title[Theoretical calculation of finite-temperature XANES]{Theoretical calculation of finite-temperature X-ray absorption fine structure: application to sodium K-edge in NaCl}

\begin{document}

\sloppy

\begin{abstract}

\end{abstract}

We present a comprehensive computational framework for reproducing the full X-ray absorption fine structure (XAFS) through quantum-chemical simulations. The near-edge region is accurately captured using an efficient implementation of time-dependent density-functional perturbation theory applied to core excitations, while \emph{ab initio} molecular dynamics provides essential sampling of core-excitation energies and interatomic distance distributions for interpreting extended X-ray absorption fine structure (EXAFS) features. Owing to the efficiency of the approach, the total spectrum can be decomposed into contributions from bulk, defective, and surface environments, which commonly coexist in experimental systems. The methodology is demonstrated for sodium at the Na K-edge in NaCl, where the predicted spectra show good agreement with experimental measurements on thin film samples. This strategy offers a practical route to generating chemically specific XAFS cross-section data for elements and species that remain challenging to characterize experimentally, thereby enabling deeper insights into materials of technological importance.

\section{Introduction}

Materials characterization plays a central role in modern science and technology, as it provides the fundamental understanding of structure-property relationships that drive innovation. By probing materials at the atomic, molecular, and macroscopic levels, characterization techniques reveal the chemical composition, electronic structure, and physical properties essential for tailoring performance.\cite{leng2013materials} From energy storage and catalysis to nanotechnology and biomedical applications, reliable characterization not only accelerates the discovery of new materials but also ensures their functionality, stability, and scalability in real-world technologies.

Experimental characterization techniques, particularly spectroscopic methods, are indispensable for unveiling the microscopic properties of materials. Spectroscopy provides direct insight into vibrational, electronic, and structural features, enabling researchers to probe bonding environments, defect states, and dynamic processes with high precision. Techniques such as infrared, Raman, UV-Vis, and X-ray spectroscopy bridge the gap between fundamental theory and practical application, offering powerful tools to monitor material behavior under operating conditions and to guide the design of next-generation functional materials.\cite{agnello2021spectroscopy,singh2021modern}

In particular, X-ray spectroscopies provide powerful insights into the electronic and structural properties of materials, making them especially valuable for the study of energy-related systems. Techniques such as X-ray absorption fine structure (XAFS) spectroscopy, which includes X-ray absorption Near Edge spectroscopy (XANES) and extended X-ray absorption fine structure (EXAFS) spectroscopy, and X-ray photoelectron spectroscopy (XPS) allow for a reliable probing of oxidation states, local coordination environments, and electronic structures with element-specific sensitivity. This enables a detailed understanding of catalytic activity, charge transport, and chemical transformations in batteries, solar cells, and other energy technologies. By capturing both static and dynamic information, X-ray spectroscopies help unravel the mechanisms that govern material performance and stability under realistic operating conditions. Recent works underscore how XAS has become a powerful probe for energy materials under realistic operating conditions. For a recent review on the application of XAFS across Li-ion, Na-ion, and Li–S batteries, see Ref.\,\citenum{fan2025operando} and the references therein. 

With the rise of sodium-based batteries as alternative to lithium-based system, many X-ray spectroscopy studies have been carried out to characterize different components of these batteries and decipher mechanisms of the underlying processes during battery operation.\cite{brennhagen2024sodiation,plotek2024sb,zhang2024advances,shipitsyn2023understanding,jeong2024pore,wu2024recent} Correct analysis and quantitative interpretation of X-ray spectroscopies require accurate key atomic fundamental parameters of sodium as well as reliable reference XAFS spectra for relevant sodium chemical species. However, the current knowledge on the fundamental parameters for sodium is limited to outdated data often derived from interpolations of neighboring elements, relying solely on theoretical calculations without experimental validation, and usually suffering from unknown or only estimated uncertainties. 

Theoretical calculations of X-ray spectroscopies provide a crucial framework for interpreting and complementing experimental measurements. By simulating core-level excitations and electronic transitions with atomistic accuracy, theory helps disentangle overlapping spectral features and assign them to specific local environments, oxidation states, or bonding motifs. Such calculations bridge the gap between raw experimental spectra and the underlying microscopic processes, enabling a more reliable identification of structure–property relationships. In energy-related materials, this combined approach is particularly powerful for tracking redox mechanisms, defect states, and degradation pathways, thereby guiding the rational design of more efficient and durable technologies. In fact, theoretical calculations may even be required to determine reference XAFS spectra for relevant but chemically highly sensitive species for which isolated experimental data within operando XAFS datasets of batteries is hard or impossible to be obtained.\cite{Zech2021}

In this work, we present practical steps towards theoretical calculation of the compound specific fine structure components, which allows for the interpretation of experimental XAFS and EXAFS data by combining quantum-chemical calculation of core-excitation energies and amplitudes and {\it ab initio} molecular dynamics simulations at finite temperatures. As a target system, we concentrate on the Na K-edge in NaCl in order to assess and benchmark the computations against experimental data. The presented computational procedure will enable us to eventually find the missing chemically specific cross-section data in existing fundamental parameters tabulations for sodium as well as for sodium species, which are relevant for Na-ion based battery materials but very hard to be addressed experimentally because of high chemical sensitivity or instability. 

\section{Methodology}

\begin{figure}
    \centering
    \includegraphics[width=0.45\linewidth]{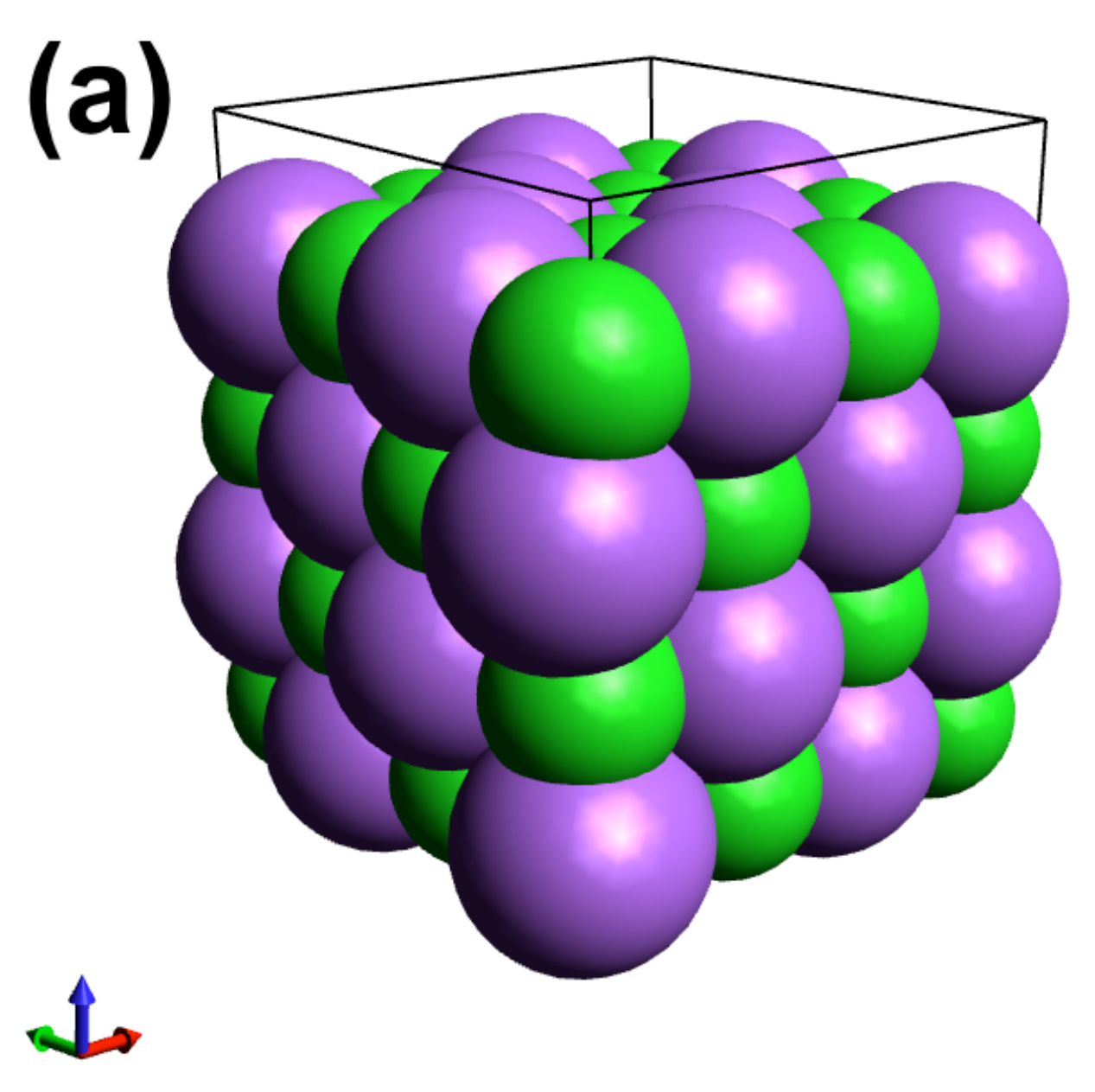}
    \includegraphics[width=0.35\linewidth]{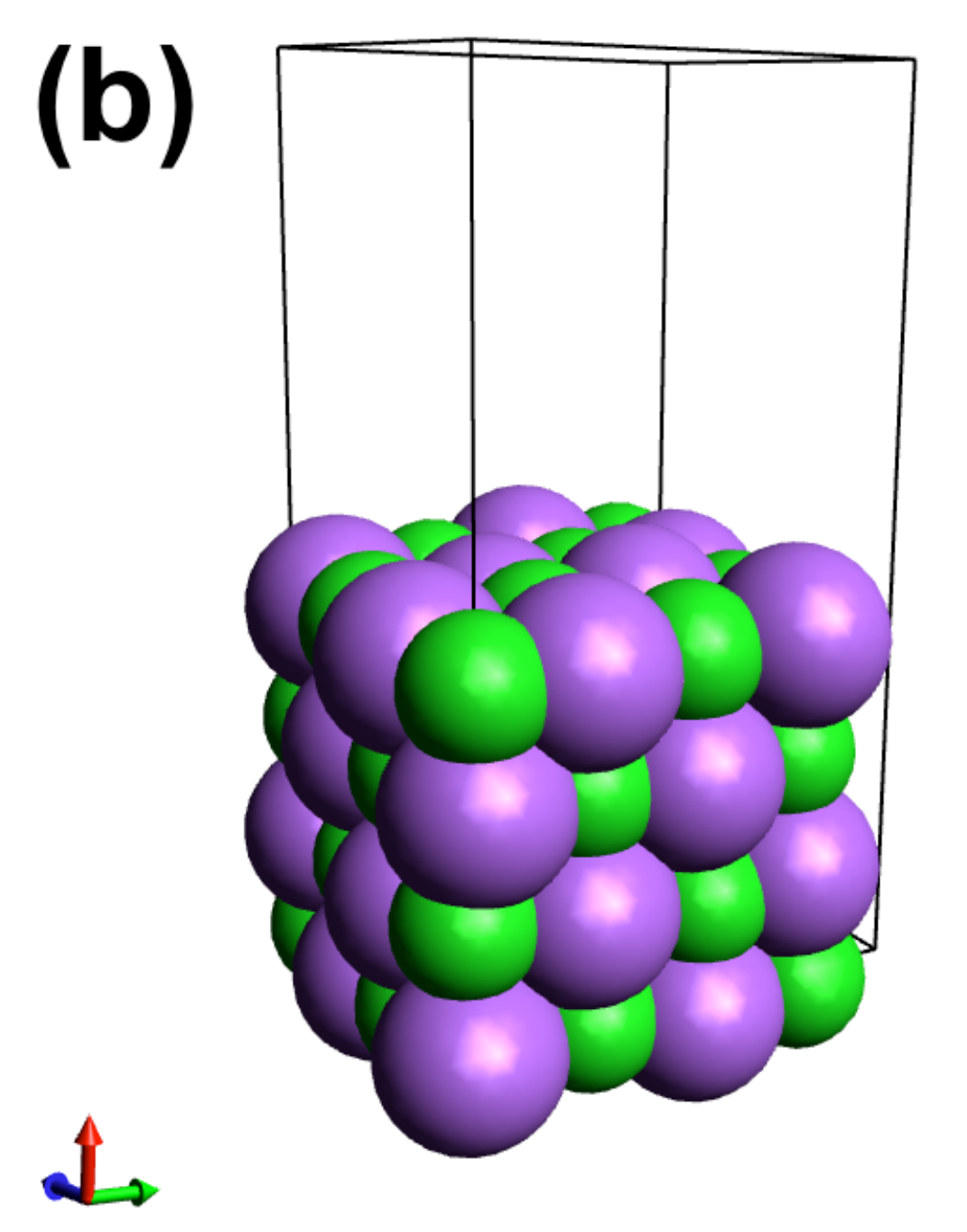}
    \caption{AIMD snapshots of the structures representing (a) NaCl bulk structure, and (b) NaCl (100) surface together with their respective unit cells considered here. Purple and green spheres denote sodium and chlorine atoms, respectively.}
    \label{fig1}
\end{figure}

The theoretical investigation have been performed for three systems, namely a perfect NaCl crystal, a defective NaCl crystal, both using a 11.176\AA$\times$11.176\AA$\times$11.176\AA\, unit cell, and a NaCl (100) surface using a 22.000\AA$\times$11.176\AA$\times$11.176\AA\, unit cell (Fig.\,\ref{fig1}). The unit cells for the perfect crystal and (100) surface contained 64 atoms, whereas two chlorine atoms were removed from the one considered for the defective sample. Periodic boundary conditions were used for all calculations. 

Finite-temperature effects have been included by performing {\it ab initio} molecular dynamics (AIMD) simulations at two different temperatures, namely $T=300$\,K and $T=400$\,K, using the CP2K software package.\cite{kuehne-2020} The AIMD simulations have been performed at electronic ground state and serve to provide a set of snapshots of each system at a finite temperature to be further used for XANES simulations. For each system, an AIMD simulation has been first carried out in the canonical ensemble for 20\,ps using CSVR thermostat.\cite{bussi2007canonical} A time step of 1\,fs was chosen for atomic coordinate propagation. The atomic forces for integrating the equation of motion during AIMD have been calculated at density-functional theory (DFT) level,\cite{hohenberg-1964,kohn-1965} where the core electrons have been effectively taken into account via Goedecker-Tetter-Hutter-type pseudopotentials,\cite{goedecker-1996,hartwigsen-1998,krack-2005} while the valence electrons have been described using the mixed Gaussian and plane wave basis set DZVP-MOLOPT-SR.\cite{vandevondele-2007} Exchange and correlation effects have been approximated using the PBE functional\cite{perdew-1996,perdew-1997} together with the semi-empirical DFT-D3 method\cite{grimme-2010} to correct for the long-range dispersion interactions. A plane-wave energy cutoff of 500\,Ry and a relative cutoff of 50\,Ry were chosen. After the equilibration phase, the atomic coordinates have been sampled in the micro canonical ensemble for an additional 20\,ps. A similar procedure involving AIMD simulations has been already adopted before for {\it ab initio} simulation of finite-temperature vibrational spectra.\cite{kiani2023characterization,kiani2025ab,partovi2023efficient,partovi2015evidence} To simulate the XANES, these trajectories have been used afterwards to calculate core-excitation energies and corresponding excitation amplitudes. 

The core excitations have been studied using time-dependent density-functional perturbation theory (TDDFPT) within Tamm-Dancoff approximation.\cite{hirata1999time} To calculate the core excitations, we have assumed that, the core and valence states only weakly couple, due to large energy differences. Following the work of Bussy and Hutter,\cite{bussy2021efficient} We have also assumed that the relaxation of electrons beyond the core region is negligible upon excitation of a core electron. Adopting the above approximations, all required 4-center 2-electron integrals involve the excited core state (here sodium 1s) and, therefore, allows for a core-specific resolution of the identity scheme.\cite{bussy2021efficient} A PBEh ($\alpha$=0.45)\cite{adamo1999toward} hybrid exchange-correlation functional has been used. All-electron basis set aug-pcseg-2\cite{jensen2015segmented} has been employed only for one Na atom in the structure while other atoms have been treated within the pseudopotential formalism to reduce the computational cost. Periodic Hartree-Fock calculations have been performed for both systems using the truncated coulomb potential\cite{guidon2009robust} together with the auxiliary density matrix method and corresponding basis set augadmm-2\cite{kumar2018accelerating} only on the excited Na atom, where core excitations from 1s level have been investigated. The core-excitation energies and oscillator strengths have been computed on reference trajectories described above and sampled every 10 fs. TDDFPT calculations have also been performed using CP2K software. The excitation energies from core level obtained from TDDFPT calculations should normally be shifted to match the experiments. This is related to the well-known self-interaction error in DFT calculations as well as to the lack of orbital relaxation upon the creation of the core hole in TDDFPT scheme adopted here. However, this can be substantially corrected for using the {\it ab initio} correction method GW2X$^*$.\cite{shigeta2001electron,bussy2021first} 

For comparison purposes, a NaCl thin film coating on a silicon nitride (SiN) membrane is used.  The NaCl coating has a nominal mass deposition of (102.5 $\pm$ 5.2) $\mu$\,g\,cm$^{-2}$, corresponding to a thickness of about 470\,nm using the tabulated bulk density of NaCl, while the carrier membrane has a nominal thickness of 1000\,nm. Both the XANES and the EXAFS of this sample at the Na K-edge have been measured in transmission mode employing the plane grating monochromator beamline of the Physikalisch-Technische Bundesanstalt at BESSYII\cite{F.Senf1998}. For reference purposes and for subtraction of the carrier membrane contribution, also a blank membrane of nominally identical thickness was used. These experiments have been carried out using an in-house developed ultra-high vacuum chamber,\cite{J.Lubeck2013} which is equipped with calibrated photo diodes used to determine and monitor the incident photon flux at each probed monochromatic photon energy. The samples were aligned to be in the center of the experimental chamber by an $x$-$y$-scanning stage. For all experiments, both the sample orientation with respect to the synchrotron radiation beam was 45\textdegree. For the transmission experiments, the incident photon energy was varied in small steps (fractions of one eV) in the vicinity of the Na-K attenuation edge and in larger steps further away from this edge. For each incident photon energy, several readings of a photo diode placed in the beam behind the sample were recorded and averaged. Further details can be found in Ref.\,\citenum{Venzke_2025}.

From the transmission data, the sample specific mass attenuation coefficients (product of the mass attenuation coefficient $\mu$, material density $\rho$ and thickness $d$) can be derived employing the Beer-Lambert law. In the case of the NaCl on SiN, it represents the sum of the sample specific mass attenuation of NaCl and SiN, respectively. Using the transmission data from the blank SiN membrane, the contribution for NaCl can then be isolated and the sample specific mass attenuation factors $\mu_{S,NaCl}(E_0)\rho d$ for NaCl can be obtained. The contribution from the Na K-shell was isolated by scaling the relevant partial subshell contributions using Ebel polynomials \cite{H.Ebel2003} into the $\mu_{S,NaCl}(E_0)\rho d$ dataset. As the relative contributions for the coherent and incoherent scattering are low in this photon energy regime, the isolated Na K-shell contribution is the sample specific photoionization cross section $\tau_{\text{Tot,NaCl}}(E_0)\rho d$.

The XAFS spectra were analyzed using the Xraylarch software package.\cite{newville2013larch} The $E_0$ value (ionization threshold) was determined using the maximum in the derivative of the XAFS spectrum. First- and second-order pre- and post-edge polynomials were used to produce a normalized spectrum. The autobk algorithm included in Larch was then used to calculate the $\chi(k)$, the EXAFS structure in $k$-space. The $k^2$-weighted EXAFS data were then Fourier-transformed to $r$-space using an FFT, with a Hanning window ranging from 2 \AA$^{-1}$ to 10 \AA$^{-1}$ with an edge width of 1 \AA$^{-1}$ being applied.

\section{Results and discussion}

\begin{figure}
    \centering
    \includegraphics[width=0.5\linewidth]{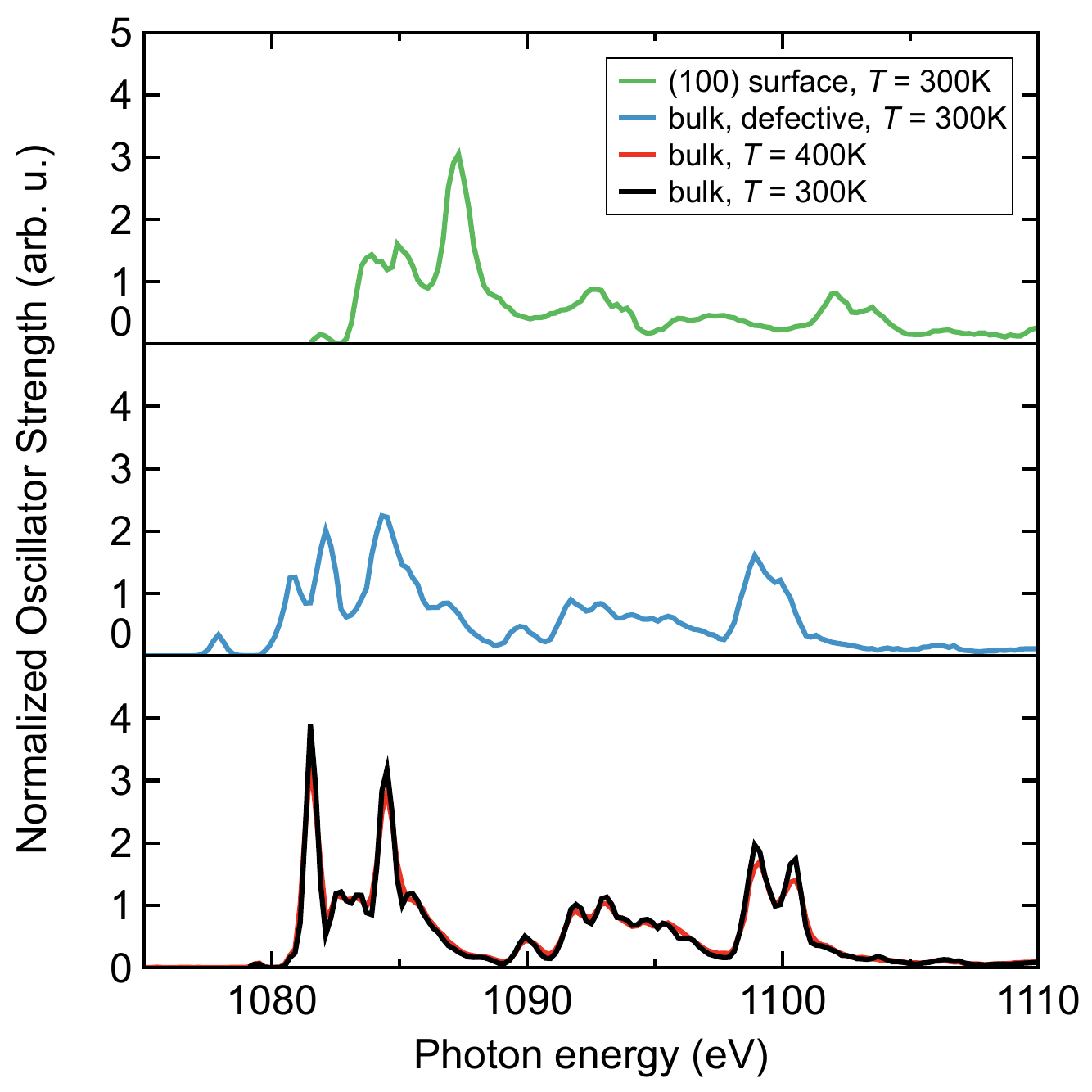}
    \caption{Computed XANES of bulk NaCl perfect crystal at $T=300$\,K (black) and $T=400$\,K (red), as well as defective sample (blue) and NaCl (100) surface (green), both at $T=300$\,K.}
    \label{fig2}
\end{figure}

The calculated XANES of the above systems are presented in Fig.\,\ref{fig2}. As seen in the lower panel, the elevated temperature results in broader peaks and overall smoother, less-featured spectrum.  
It should be emphasized again that here the effect of finite temperature on the spectrum has been included only through computation of oscillator strengths for core excitations averaged over an AIMD trajectory. A pre-edge activity is also seen in the XANES of the perfect crystals at both temperatures. Main features above 1090\,eV are also observed in the bulk defective system (blue). Although the photon energies corresponding to the resonances close to the threshold (1080-1085\,eV) are also seen in the defective system, but their relative intensities differ significantly. Additionally, a small peak is observed at lower energies  in the XANES for the defective sample, $\sim$2\,eV lower than the pre-edge peak in the perfect systems. The spectrum for the (100) surface exhibits the substantial difference as compared to the spectra of the perfect and defective bulk systems. Apart from the resonances in 1080-1082\,eV range becoming considerably insignificant, a strong peak is observed  at around 1087\,eV and a broader one above 1100\,eV.

\begin{figure}
    \centering
    \includegraphics[width=0.95\textwidth]{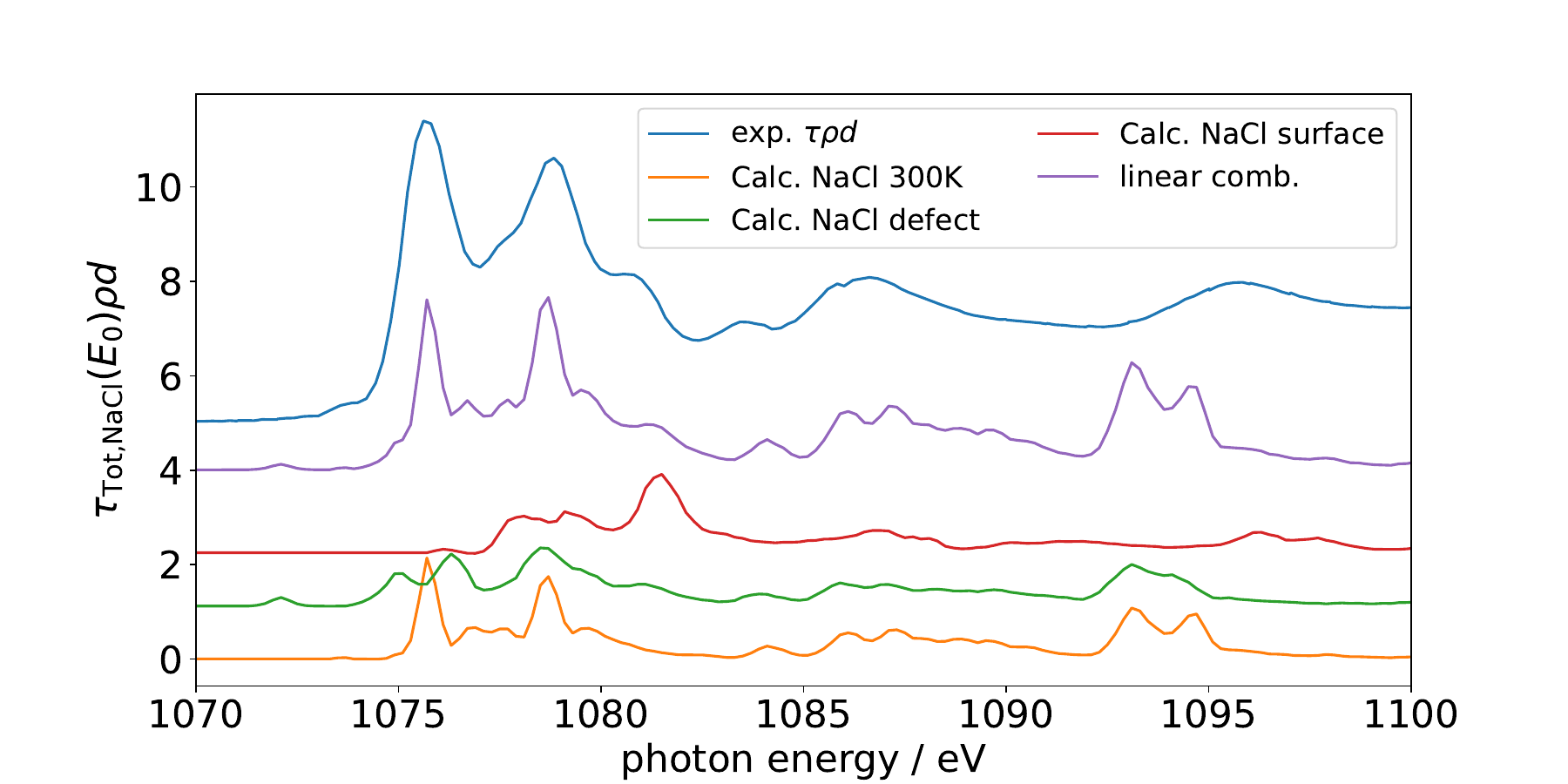}
    \caption{Calculated XANES for the perfect NaCl crystal (orange), defective crystal (green), and NaCl (100) surface (red). Experimental spectrum measured for a thin film sample is also shown (blue). Purple curve represents a linear combination of the computed spectra with equal weights. The spectra were shifted vertically for clarity.}
    \label{fig3}
\end{figure}

In order to incorporate the effects from different systems to an X-ray near edge spectrum comparable to and experimental XAFS spectrum of a thin film sample (which presumably encompasses perfect and defective NaCl crystallites as well as different crystal surfaces) we have computed a ``total'' spectrum as a linear combination of the individual spectra presented in Fig.\,\ref{fig2}. Figure\,\ref{fig3} shows the calculated XANES for the perfect NaCl crystal (orange), defective crystal (green), and NaCl (100) surface (red), as well as the experimental XANES measured for a NaCl thin film sample (blue). Also shown in Fig.\,\ref{fig3} is a linear combination of computed spectra, with manually selected weights (purple, 63 \% of the perfect NaCl crystal, 26 \% of defective crystal and 11 \% of the NaCl (100) surface).
All the computed spectra are rigidly shifted in energy to match the energy corresponding to the first experimentally observed resonance. %\sout{Additionally, the tabulated mass attenuation coefficients of NaCl\cite{schoonjans2011xraylib} is added to the calculated overall spectrum (purple) for the relative intensities to be more comparable to experiments.} 
%\textcolor{red}{Argument on the fact that the linear combination produces a spectrum which is agrees very well with the experimental measurement in the near edge region. For the extended region, we will put the modified figure from Athena and compare it to the RDFs in Fig. 1...}
Without the rigid shift, there is about 5\,eV difference between the experimentally measured and theoretically predicted energy of the first resonance. The employed GW2X$^*$ considerably decreases the difference between the calculated and measured energies. The observed discrepancy is due to the self-interaction error in the DFT calculations and the lack of orbital relaxation upon the creation of the core hole.\cite{bussy2021first} The relative intensities and peak positions are well reproduced by the purple spectrum. The agreement between the theory and experiment is remarkable up to 1090\,eV. The discrepancy between the purple and blue curves is more profound at around and above 1095\,eV, which might be an indicator that the contribution of the NaCl surfaces (red) to the overall spectrum could be more substantial. However, it should be noted that the actual crystal structure of the NaCl thin film on the SiN carrier membrane is likely to be more complex.\\

\begin{figure}
    \centering
    \includegraphics[width=0.95\textwidth]{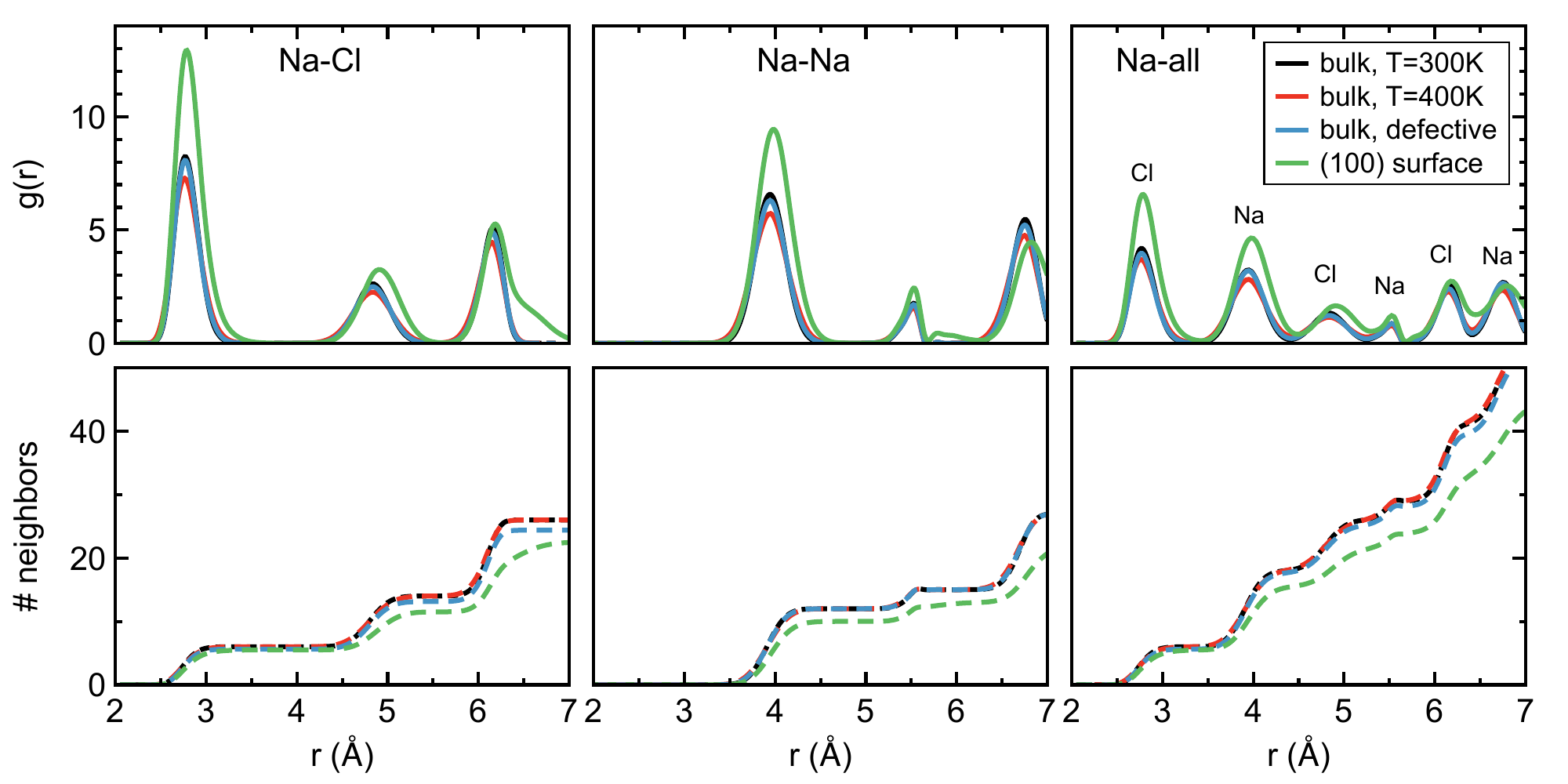}
    \caption{Radial distribution functions (upper panels) of Na-Cl (left), Na-Na (middle), and Na-all (right) in different systems, namely perfect crystal at $T=300$K (black), perfect crystal at $T=400$K (red), defective crystal (blue), and NaCl (100) surface (green). Please note  the different normalization factors for $g(r)$ in each system. The panels on the lower row show the corresponding number of neighboring atoms from a typical sodium atom.}
    \label{fig4}
\end{figure}

The calculated radial distribution functions, $g(r)$, are shown on upper panels of Fig.\,\ref{fig4}, while the lower panels show the number of neighbors of typical Na reference atom.
While the radial distribution functions and the number of neighboring atoms in the perfect (both at $T=300$\,K and $T=400$\,K) and defective samples slightly differ from one another, the most profound difference is observed in the case of NaCl (100) surface. The nearest neighbor distance from a reference Na atom (Cl) in all the systems is found to be around $2.75$\AA, while the second nearest neighbor (Na) is found at about $3.95$\AA. The thirst nearest neighbor (Cl) is slightly shifted from $4.85$\AA\, in the perfect and defective samples to $\sim$$2.95$\AA\, in the case of (100) surface. The upper-right panel in Fig.\,\ref{fig4} reflects this observation. 

\begin{figure}
    \centering
    \includegraphics[width=0.45\linewidth]{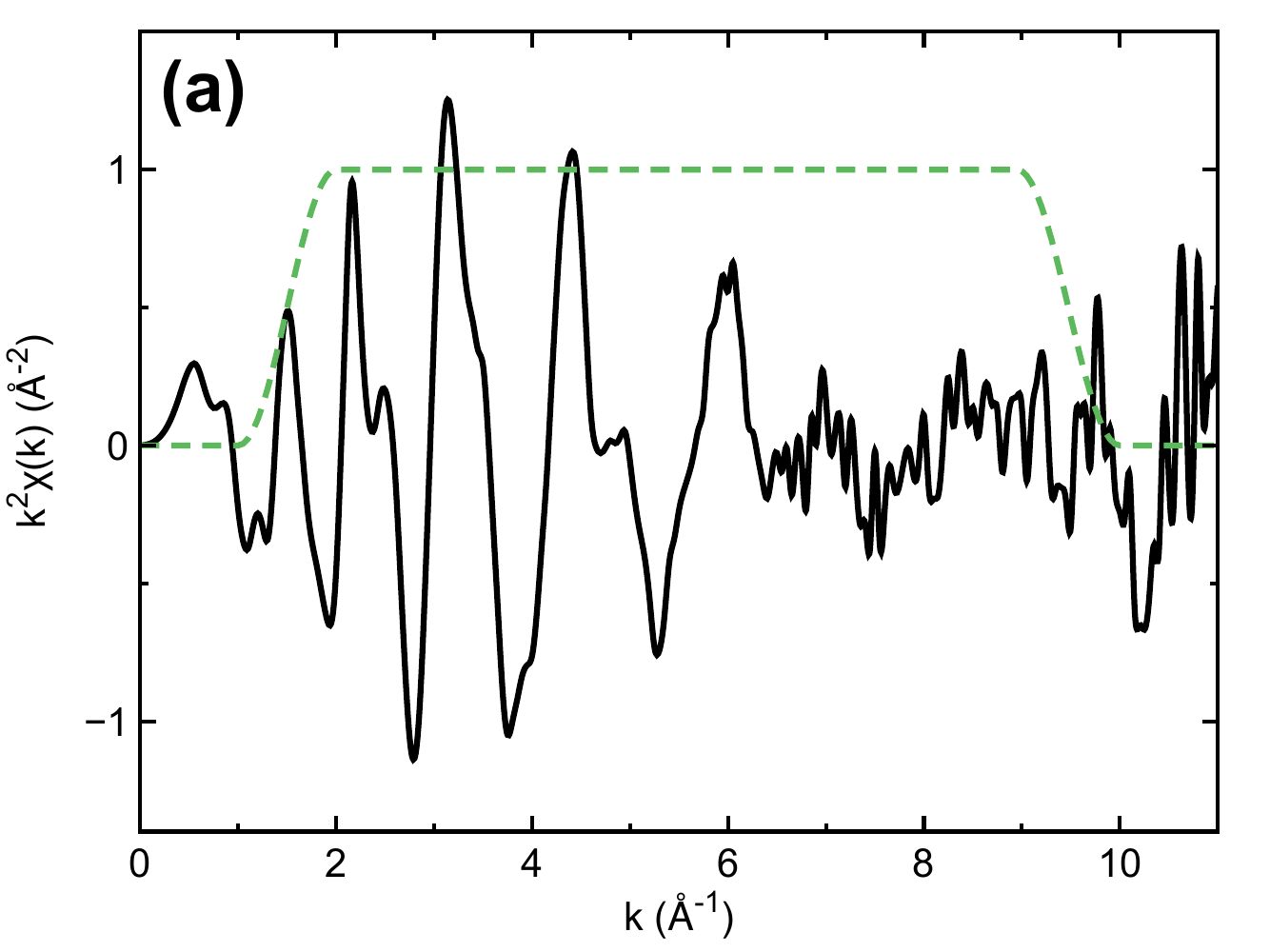}
    \includegraphics[width=0.45\linewidth]{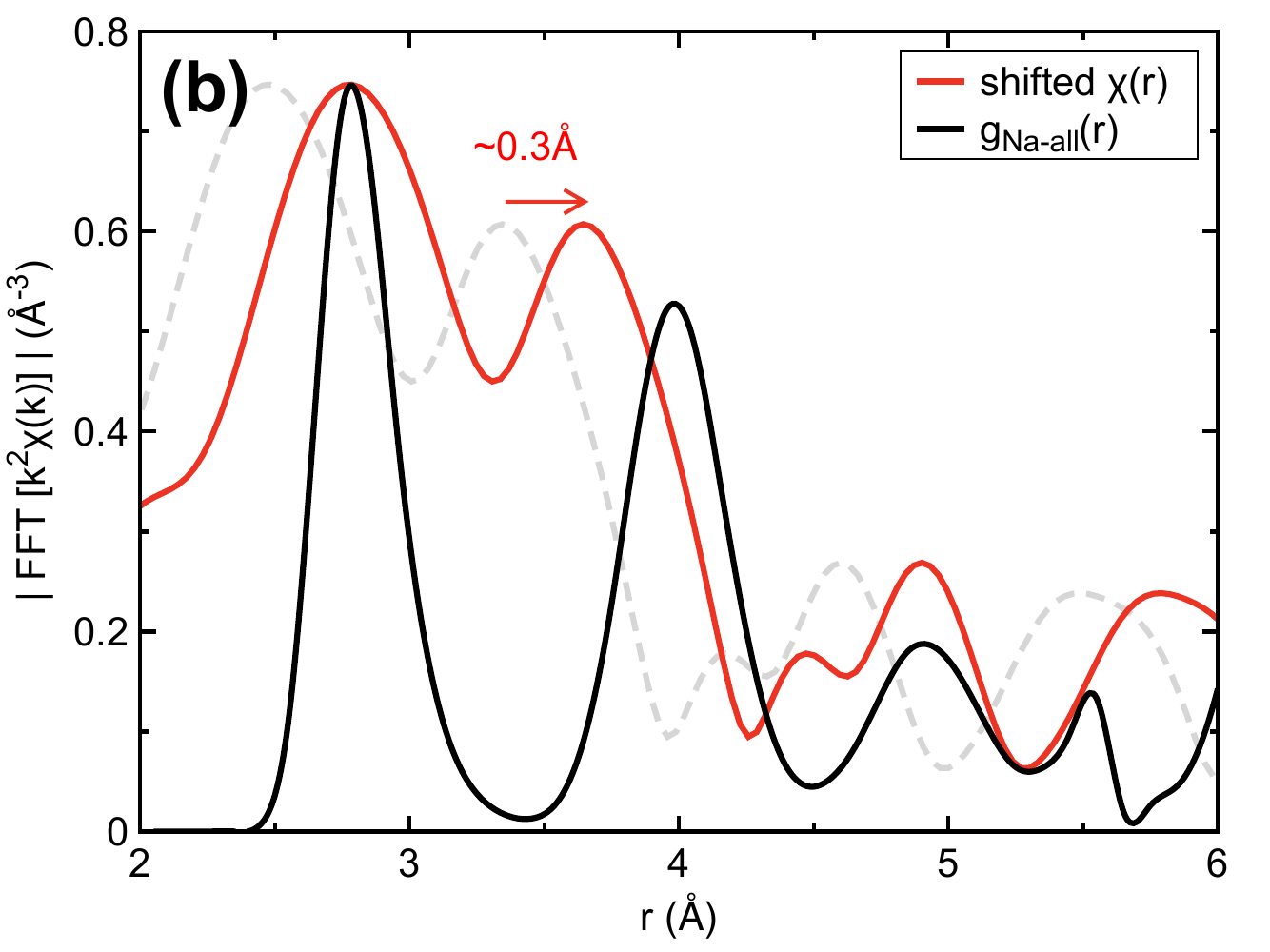}
    \caption{(a) EXAFS data (black) together with used Hanning windows (green). (b) Magnitude of the non-phase-shift-corrected FFT of the EXAFS data (light gray) together with the same FFT data rigidly shifted by $\sim$0.3\,eV (red). Also shown in black is a scaled radial distribution function, $g_{\rm Na-all} (r)$ obtained from the AIMD simulations on the defective crystal system (Fig.\,\ref{fig4}). For a better comparison, the rigid shift and the scaling factor were chosen so that the first peaks in the red and black curve coincide in both position and intensity.}
    \label{fig5}
\end{figure}

The EXAFS data together with the applied Hanning window is shown in Fig.\,\ref{fig5}(a), while The magnitude of the non-phase-shift-corrected FFT obtained from the experimental EXAFS data in $r$-space is shown in Fig.\,\ref{fig5}(b) in light gray.
The discrepancy between the experimental data and the calculations is due to a phase-shift in the scattering process. The same FFT data with a rigid shift of $\sim$0.3\,eV is shown in red. The black curve in Fig.\,\ref{fig5}(b) is a scaled radial distribution function, $g_{\rm Na-all} (r)$ obtained from the AIMD simulations on the defective crystal system (blue curve in upper-right panel of Fig.\,\ref{fig4}). For a better comparison, the rigid shift and the scaling factor were chosen so that the first peaks in the red and black curve coincide in both position and intensity. Although the agreement between the experimental and theoretical results seems acceptable, the main difference appears to occur at around 3.65\,\AA\, and about 4.50\,\AA. This could be related to considerable difference in the structure of thin film samples used in the measurements and relatively simple systems considered in computations. Additionally, the discrepancy at above 7\,\AA\, could be related to relatively small unit cell size in the calculations.

\section{Conclusions}

In this work, we have shown that the complete X-ray absorption fine structure (XAFS) can be reliably reproduced using state-of-the-art quantum-chemical methods. The near-edge spectral features are accurately described through an efficient implementation of time-dependent density-functional perturbation theory applied to core electrons, while \emph{ab initio} molecular dynamics simulations prove essential both for sampling core-excitation energies and for providing radial distribution functions that enable a direct interpretation of the extended X-ray absorption fine structure (EXAFS). The combined approach has been successfully applied to the Na K-edge in NaCl, yielding theoretical spectra in good agreement with experimental measurements on thin-film samples. Thanks to the efficiency of the employed methodology, we were further able to disentangle the spectral contributions originating from bulk crystals, lattice defects, and surfaces-structural motifs that are typically present in experimentally studied samples. Altogether, this computational framework provides a robust pathway for generating chemically specific XAFS cross sections of elements and compounds that remain difficult or even impossible to characterize experimentally.

\section*{Author contributions}
P.H. and P.P.-A contributed to Conceptualization, Funding acquisition, Formal analysis, Investigation, Methodology, Visualization, Validation, and Writing (original draft, review \& editing); Y.K. contributed to Formal analysis, Methodology, Validation, and Writing (review \& editing).     

\section*{Conflicts of interest}
There are no conflicts to declare.

%\begin{datastatement}
%Data supporting this article have been included as part of the Supplementary Information. Data for this article, including relaxed atomic coordinates of the studied systems and diffusion paths (.xyz files), as well as energy profiles for the latter (.txt files) are available at Zenodo at https://doi.org/10.XXXX/zenodo.XXXXXXX.
%\end{datastatement}

\begin{acknowledgement}
P.P.-A. gratefully acknowledges DFG funding via projects 420536636 and 446879138, as well as the computing time made available on the high-performance computer at the NHR Center of TU Dresden via the project `p\_oligothiophenes'. Y.K. thanks the Max Planck Society for funding. In addition, the work was partially supported by the project 14ACMOS (grant agreement number 101096772), which is funded by the Chips Joint Undertaking and its members, including the top-up funding of Belgium and the Netherlands. Parts of this research have been supported by Hi-Acts, an innovation platform under the grant of the Helmholtz Association HGF within the project BATIX.
\end{acknowledgement}

%\begin{suppinfo}
%
%\end{suppinfo}

\bibliography{refs.bib}

\end{document}